\documentclass[aps,prl,twocolumn,amsmath,amssymb,nofootinbib,showpacs,superscriptaddress]{revtex4-1}
\usepackage[english]{babel}
\usepackage{latexsym}
\usepackage{graphics}
\usepackage{graphicx}
\usepackage{epsfig}
\usepackage{color}
\usepackage{bm}
\usepackage{amsmath}
\usepackage{amssymb}
\usepackage{amsthm}
\usepackage{dcolumn}
\usepackage{bm}
\usepackage{float}
\usepackage{hyperref}
\usepackage{color}
\usepackage{epstopdf}
\usepackage[svgnames]{xcolor}
\usepackage{enumerate}
\usepackage{braket}
\usepackage[normalem]{ulem}
\hypersetup{hidelinks,colorlinks=true,allcolors=DarkBlue}

\theoremstyle{remark}

\begin{document}

\title{Experimental quantum homodyne tomography via machine learning}

\author{E.S. Tiunov}\email{These authors contributed equally to this work.}
\affiliation{Russian Quantum Center, Skolkovo, Moscow 143025, Russia}
\affiliation{Moscow Institute of Physics and Technology, Dolgoprudny, Moscow Region 141700, Russia} 

\author{V.V. Tiunova (Vyborova) }\email{These authors contributed equally to this work.}
\affiliation{Russian Quantum Center, Skolkovo, Moscow 143025, Russia}

\author{A.E. Ulanov}
\affiliation{Russian Quantum Center, Skolkovo, Moscow 143025, Russia}

\author{A.I. Lvovsky}\email{alex.lvovsky@physics.ox.ac.uk}
\affiliation{Russian Quantum Center, Skolkovo, Moscow 143025, Russia}
\affiliation{Department of Physics, University of Oxford, Oxford OX1 3PG, UK}

\author{A.K. Fedorov}\email{akf@rqc.ru}
\affiliation{Russian Quantum Center, Skolkovo, Moscow 143025, Russia} 
\affiliation{Moscow Institute of Physics and Technology, Dolgoprudny, Moscow Region 141700, Russia}

\date{\today}
\begin{abstract}
Complete characterization of states and processes that occur within quantum devices is crucial for understanding and testing their potential to outperform classical technologies for communications and computing.
However, solving this task with current state-of-the-art techniques becomes unwieldy for large and complex quantum systems. 
Here we realize and experimentally demonstrate a method for complete characterization of a quantum harmonic oscillator based on an artificial neural network known as the restricted Boltzmann machine.
We apply the method to optical homodyne tomography and show it to allow full estimation of quantum states based on a smaller amount of experimental data compared to state-of-the-art methods. 
We link this advantage to reduced overfitting.
Although our experiment is in the optical domain, our method provides a way of exploring quantum resources in a broad class of large-scale physical systems, 
such as superconducting circuits, atomic and molecular ensembles, and optomechanical systems.
\end{abstract}

\maketitle
\section{Introduction}
Exploiting the full potential of quantum technologies involves the challenge of `quantum volume': 
keeping a high degree of control over a complex many-body quantum system in spite of its growing size~\cite{Smolin2017}. 
This important challenge concerns, in particular, methods for complete characterization of quantum states and processes.
Quantum state tomography (QST), the reconstruction of quantum states from measurement statistics in multiple bases~\cite{Leonhardt1997,Lvovsky2009}, 
is routinely performed in quantum physics experiments of various nature. 
Nevertheless, because the number of parameters describing a state of a quantum system grows exponentially with its size, 
tomography becomes increasingly demanding in application to large-scale quantum systems that are now engineered in experiments with ultracold atoms~\cite{Browaeys2016,Lukin2016,Lukin2017,Browaeys2018},  
ions~\cite{Monroe2017,Blatt2018,Blatt2019}, superconducting devices~\cite{Martinis2018}, and quantum light~\cite{QuantumLight}. 

This problem manifests itself in two aspects. 
First, full quantum tomography of multi-dimensional quantum systems requires large portions of data, which are typically difficult to acquire experimentally. 
Second, even if such data are available, they are quite difficult to process with reasonable computational resources. 
Fortunately, it often happens that the physical setting being studied imposes certain \emph{a priori} restrictions on the quantum states that can be prepared in it. 
As a result, the states can be described using a set of parameters that grows polynomially, rather than exponentially, with the size of the system. 
This observation gave rise to alternative approaches such as permutationally invariant tomography~\cite{Toth2010}, quantum compressed sensing~\cite{Gross2010}, and tensor networks~\cite{Cramer2010,Lanyon2017,Carrasquilla2019}. 
Each of these approaches makes particular assumptions about the physical restrictions imposed upon the state in question. 

In the absence of knowledge about the physics of the system, one can use a universal approach based on generative artificial neural networks. 
Generally, neural networks are known to be capable of finding the best fit to arbitrarily complex data patterns with a limited number of parameters available~\cite{Cybenko1989}.
In the context of quantum physics, this capability has been exploited in the context of neural networks known as the restricted Boltzmann machine (RBM).  
Such a neural net is proven to be a universal approximator for any discrete distribution~\cite{Bengio2008}.
RBMs are capable to encode the information about exponentially many terms of a quantum state in a polynomial number of units~\cite{Troyer2017}. 
This feature makes RBMs attractive for a variety of quantum variational optimization problems~\cite{Melko2019}, which require finding a quantum state that best satisfies a certain criterion. 
Examples of such problems, in addition to quantum tomography~\cite{Troyer2018}, 
include searching ground states of Hamiltonians in quantum chemistry tasks~\cite{Xia2018}, investigating tensor network states~\cite{Chen_PRB} and topological states~\cite{Lu2019}, 
and simulating open quantum many-body systems~\cite{Schuld2019,Yoshioka2019,Nagy2019,Hartmann2019,Vicentini2019}.

In the original theoretical proposal~\cite{Troyer2018}, RBM-based QST has been applied to simulated pure states of interacting many-qubit systems. 
A subsequent work~\cite{Torlai2018} has generalized this approach to mixed states and applied it to perform QST of a two-qubit system associated with a polarization-entangled photon pair. 
Very recently, the method was used in application to an experimental Rydberg-atom simulator with eight and nine atoms, using a pure-state, constant-phase approximation and measurements in a single basis~\cite{Torlai2019}.
Neural network techniques in the context of QST were also employed, albeit in a very different setting, to pre-process the data, thereby reducing the effect of state preparation and measurement errors~\cite{Palmieri2019}.

However, all existing work on the subject has been applied to sets of natural qubits, such as fermion spins. 
This excludes a large class of `continuous-variable' physical systems whose Hamiltonian is identical to that of the harmonic oscillator. 
These include light, superconducting cavities, atomic and molecular ensembles, and optomechanical arrangements. 
Many of these systems are promising candidates for quantum information processing \cite{Asavanant2019,Larsen2019} and hence the challenge of `quantum volume' applies to them to the full extent. 
This necessitates the extension of neural-network QST methods to these systems.

Here we fill this gap by applying the RBM to homodyne tomography of optical states, in which measurements of electromagnetic field quadratures at various phases are performed to reconstruct the state of light in a given mode~\cite{Lvovsky2009}. 
We verify our method on experimental data for the cases of optical Schr{\"o}dinger's cat states and arbitrary Fock-state superpositions up to the two-photon level, where we obtain high quality of quantum state reconstruction. 
We perform the universality test for our method via the reconstruction of randomly-generated states.
We also consider the application of our methods to other relevant quantum states, such as Gottesman-Kitaev-Preskill states~\cite{Gottesman2001} and squeezed-displaced vacuum.
The approach generally outperforms standard maximum-likelihood based methods~\cite{Lvovsky2004}, which, as we demonstrate, is deeply linked with reduced overfitting.  
To our knowledge, this is the first application of neural networks in continuous-variable quantum setting.

\begin{figure}[t!]
	\begin{center}
	\includegraphics[width=0.8\columnwidth]{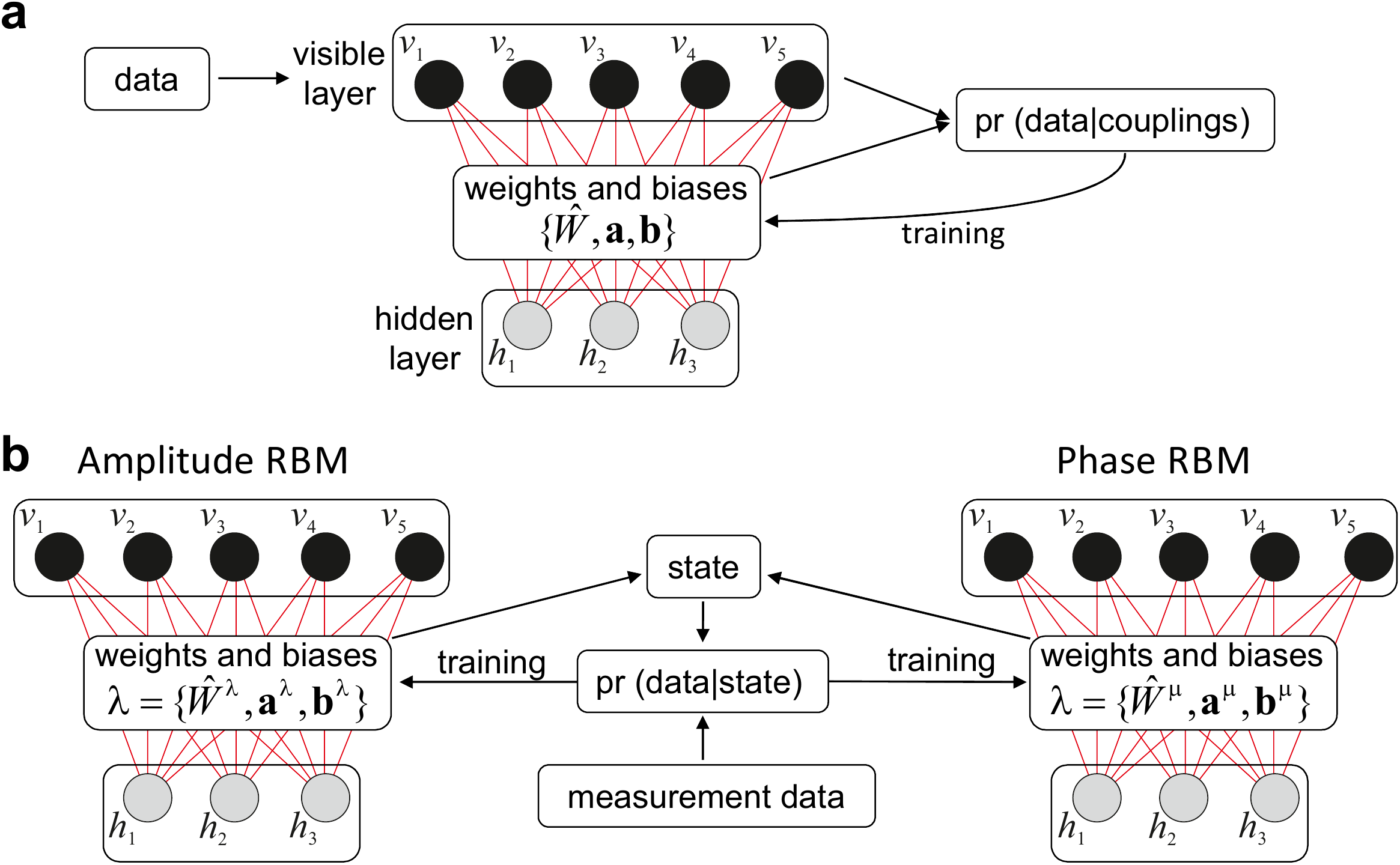}
	\end{center}
	\vskip -4mm
	\caption{Architecture of restricted Boltzmann machines for classical pattern recognition tasks (a) and quantum tomography (b).}
	\label{fig:architecture}
\end{figure}

\section{Neural network tomography}

An RBM is a neural net containing two layers, visible and hidden, with all-to-all connections between the neurons in different layers and none inside each layer [hence the term `restricted', see Fig.~\ref{fig:architecture}(a)]. 
The neurons can take on binary values $\{0,1\}$. 
Any set of neuron values, defined by binary vectors $\mathbf v$ and  $\mathbf h$, is associated with the Boltzmann probability
\begin{equation}
\label{Boltzmann_joint}	
	p(\mathbf v,\mathbf h) = \frac{1}{Z} e^{-E(\mathbf v,\mathbf h)},
\end{equation}
where $Z$ is the partition function and $E(\mathbf v,\mathbf h)$ is the Ising-type energy functional
\begin{equation}
\label{Energy}
	E(\mathbf v,\mathbf h) = -\mathbf v^T \hat W \mathbf h - \mathbf a^T \mathbf v - \mathbf b^T \mathbf h
\end{equation}
with $\hat W$, $\mathbf a$, $\mathbf b$ are the network parameters: weights and biases, respectively. 
The conventional RBM is trained to find the parameter set that maximizes the product of  marginal distributions,
\begin{equation}\label{pv}
	p(\mathbf v) = \sum_{\mathbf h} p(\mathbf v,\mathbf h),
\end{equation}
over the training set $\{\mathbf{v}\}$, i.e. $\prod_{\{\mathbf{v}\}}p(\mathbf v)  $. 
The RBM trained in this way will produce similarly low energy values for test inputs that are similar to elements of the training set, which is useful for pattern recognition \cite{Larochelle2008}. 
Furthermore, by sampling high-probability visible layer vectors, one can use the RBM as a generative neural network \cite{hu2016deep}. 

In the classical case, the data (such as the pattern to be recognized) are fed to the RBM through the visible layer. 
Doing so for quantum tomography would be unimaginable because there are infinitely many quantum states and even more possible measurement data sets. 
On the other hand, we can take advantage of our \emph{a priori} knowledge of the connection between quantum states and the measurement probabilities associated with different bases. 

These important differences dictate a different way that RBMs can be applied for quantum optimization problems. 
Here we utilize the RBMs to define an \emph{Ansatz} expression for the quantum state $\ket\Psi$, which we wish to reconstruct.  
The neural network parameters are then used as the variational parameters of that Ansatz. 
We calculate the likelihood function (probability of having acquired the present experimental data set given $\ket\Psi$) using the knowledge of quantum mechanics, and optimize the parameters, and therefore $\ket\Psi$, to maximize that likelihood. 
The visible layer no longer plays the role of the container for the data, but only serves to \emph{index} the basis of the Hilbert space: each possible configuration  $\mathbf v$ of the visible layer is associated with one and only one basis element $\ket{\mathbf v}$. 

The Carleo and Troyer Ansatz  \cite{Troyer2017}, which we utilize here, uses two RBMs of identical architectures [Fig.~\ref{fig:architecture}(b)], with the parameter sets $\lambda=\{\hat W^{\lambda},\mathbf a^{\lambda},\mathbf b^{\lambda}\}$ and $\mu=\{\hat W^{\mu},\mathbf a^{\mu},\mathbf b^{\mu}\}$
to express, respectively, the amplitudes and phases of the state's decomposition into this basis:
\begin{equation}\label{pure}
	\ket{\Psi} = \sum_{\mathbf v} \sqrt{p_{\mathbf v}} e^{i\phi_{\mathbf v}/2}\ket{\mathbf v},
\end{equation}
where 
\begin{eqnarray}\label{ampphase}
p_{\mathbf v} = \frac{1}{Z_\lambda} \sum_{\mathbf h} e^{-E^{\lambda}({\mathbf v},{\mathbf h})}, \quad \phi_{\mathbf v}  = \text{log}  \sum_{\mathbf h} e^{-E^{\mu}({\mathbf v},{\mathbf h})}
\end{eqnarray}
and $E^{\lambda,\mu}$ are defined by Eq.~\eqref{Energy} for the two corresponding RBMs. 
We note that the partition function $Z$ is present only in the expression for the amplitudes, but not phases, because the phases have no normalization requirement. 
The logarithm is included in the phase for mathematical convenience. 

\begin{figure*}[t!]
	\includegraphics[width=1\linewidth]{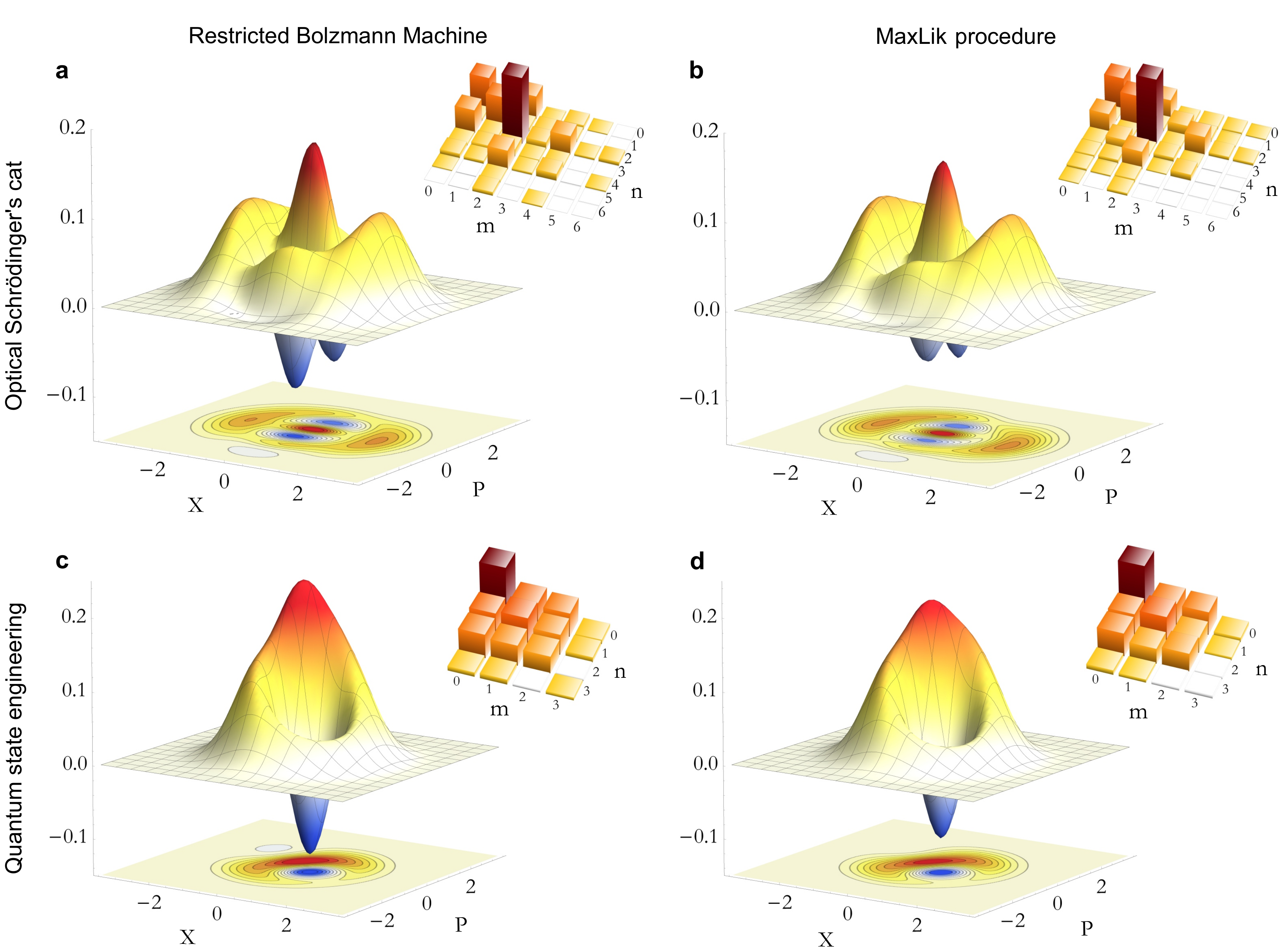}
	\caption{Experimentally reconstructed Wigner functions and density matrices for optical Schr{\"o}dinger's cats (a-b) and engineered Fock superpositions up to the two-photon level (c-d) using neural network quantum tomography (a,c) and MaxLik (b,d). 
		The relative fidelity of the two reconstructed states is about 0.998 in both cases after efficiency correction.}
	\label{exptFig}
\end{figure*}

In optical homodyne tomography, the basis traditionally used for state reconstruction is the Fock basis, bounded from above by some cut-off photon number $N_{\rm ph}$. 
Because an RBM with the visible layer of size $m$ can represent a Hilbert space of dimension $2^m$, the natural choice is to construct the reconstruction basis from photon number states ${\{\ket 0,\ldots,\ket{N_{\rm ph}=2^m-1}\}}$. 
The basis is then encoded in the visible layer in a straightforward fashion, for example, for $m=2$ :
\begin{eqnarray*}
	\ket 0\to
	\begin{pmatrix}
		0 \\ 0  
	\end{pmatrix}
	\quad
	\ket 1\to  	
	\begin{pmatrix}
		0 \\ 1  
	\end{pmatrix}
	\quad
	\ket 2\to 	
	\begin{pmatrix}
		1 \\ 0  
	\end{pmatrix}
	\quad
	\ket 3\to	
	\begin{pmatrix}
		1 \\ 1  
	\end{pmatrix}.
\end{eqnarray*}
The tomography experiment consists in measuring the continuous electromagnetic field quadrature samples $X$ on multiple copies of the state $\ket\Psi$ at various phases $\theta$. 
The log-likelihood functional is then as follows:
\begin{equation}\label{LogLik}
\Xi=\sum_j \log \,\langle \theta_j,X_j|\hat\rho|\theta_j, X_j\rangle,
\end{equation}
where $\hat\rho=\ket\Psi\bra\Psi$ is the density matrix, $j$ enumerates measurement outcomes. 
This is a differentiable function of the RBM parameters, defined through Eqs.~\eqref{Energy}, \eqref{pure}, and \eqref{ampphase}. 
These parameters can be therefore optimized using gradient descent to maximize the log-likelihood.

A general quantum tomography method must be able to work with not only pure states, but also with mixed ones. 
The method above is readily generalized to mixed states by means of purification: introducing an ancillary ``environment" Hilbert space, whose dimension is equal to that of the Hilbert space of interest. 
The mixed state that needs to be reconstructed can then be written as a partial trace 
\begin{equation}\label{TrE}
\hat\rho = {\rm Tr}_E \left( \ket{\Psi_{SE}}\bra{\Psi_{SE}} \right)
\end{equation}  
where the pure state $\ket{\Psi_{SE}}$ is a vector of the tensor product Hilbert space comprising the system and the environment and can be reconstructed from the experimental data as described above (see Supplementary for details). 
We note that, although the dimension of the tensor product space is the square of the dimension of the system, the number of visible units needed to represent that space is only twice as large as that for the system alone.

We emphasize again the difference between the RBM approach to state reconstruction and the conventional quantum expectation-maximization (MaxLik) technique \cite{Lvovsky2004,Hradil2004}. 
In both cases, we optimize the parameters of the state to maximize the likelihood functional \eqref{LogLik}.
However, in the standard approach, all elements of the density matrix are being optimized, which corresponds to the number of parameters equal to the dimension of the Hilbert space squared. 
Within the RBM Ansatz, on the other hand, the number of parameters is on the scale of the product of the number of visible and hidden units, i.e.~scales logarithmically with the Hilbert space dimension. 
As discussed previously, this is of great advantage when this dimension is large. 
Although reducing the number of parameters does restrict the set of states that can be expressed by the RBM Ansatz, 
we found it to be sufficient to adequately represent the states observed in homodyne tomography experiments.

We test our approach on two sets of experimental data. 
The first set corresponds to an optical analog of Schr{\"o}dinger's cat, i.e.~the superposition of two opposite-amplitude coherent states. 
The data have been taken from the experiment~\cite{Sychev2017} and correspond to the cat state of amplitude $\alpha=1.85$ squeezed by 3 dB along the quadrature axis. 
The second data set was obtained in an experiment on engineering arbitrary superpositions of Fock states $a_0\ket{0}+a_1\ket{1}+a_2\ket{2}$ with the amplitude ratio $a_0:a_1:a_2\sim-0.76:0.49:0.42$~\cite{Bimbard2010}. 
We compare our reconstruction results with standard iterative MaxLik algorithm with efficiency correction. For both methods, we obtain Wigner functions and density matrices of the reconstructed states (Fig.~\ref{exptFig}). 

For the reconstruction of the cat state, we used the cut-off photon number of $N_{\rm ph}=7$ (i.e.~$m=3$), which corresponds to the amplitude and phase RBMs containing $2m=6$ visible units each. 
Additionally, each RBM contained 8 hidden units. The reconstruction featured correction for 62\% detection efficiency (see Supplementary). 
For the Fock state superposition, each RBM had 4 visible units, 4 hidden units,  $N_{\rm ph}=3$ ($m=2$) and the efficiency correction 55\%.
As we see in Fig.~\ref{exptFig}, both methods resulted in similar reconstructed states, with the relative fidelity about 0.998 in both cases. 
In the Supplementary, we present the reconstruction from the same experimental data but without efficiency correction.

\begin{figure}[t!]
	\begin{center}
	\includegraphics[width=0.8\linewidth]{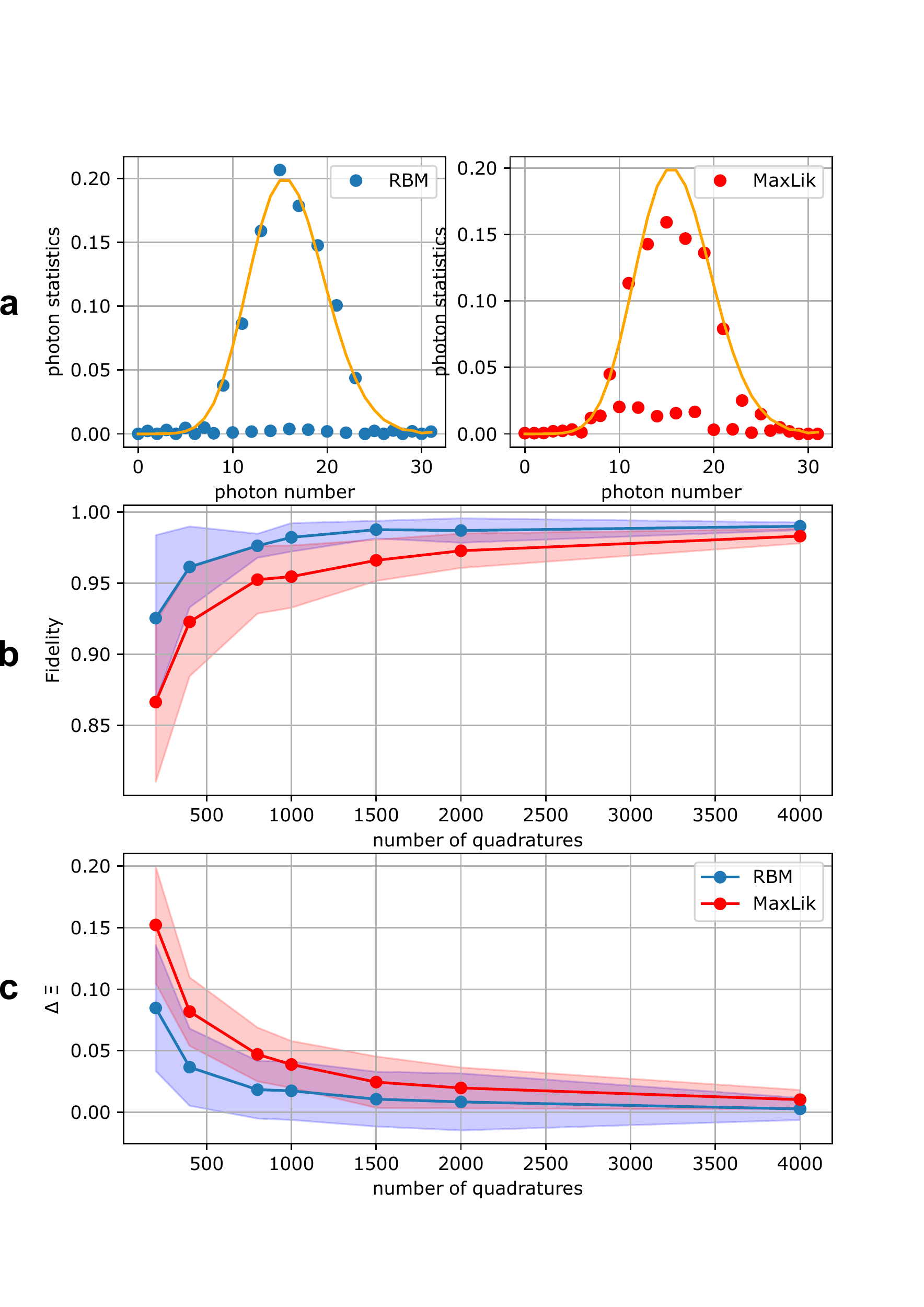}
	\end{center}
	\vskip - 12mm
	\caption{Reconstruction of the cat state $\ket\alpha-\ket{-\alpha}$ with $\alpha = 4$ from bootstrapped data. 
		a) Photon statistics for the state reconstructed from 800 quadratures using MaxLik (right) and RBMs (left). 
		b) Reconstruction fidelity as a function of   number of quadrature measurements. Each point is averaged over multiple datasets of the same size. Shaded regions show the standard deviation.
		c) Cross-validation log-likelihood difference (see text). 
		Higher values correspond to more significant overfitting. }	
	\label{fig:Comparison}
\end{figure}

\section{Effects of overfitting}
Our next goal is to compare the performance of the RBM approach to MaxLik. 
Using \emph{bona fide} experimental data is suboptimal for this purpose because it is not known what ``true" state they correspond to, and hence we cannot tell which method gives better reconstruction.

Therefore we generate a simulated quadrature data set corresponding to the Schr{\"o}dinger's cat states $\ket\alpha-\ket{-\alpha}$ with $\alpha=4$, reconstruct the state from this set and compare it to the original. 
The RBM reconstruction was performed without assuming the state to be pure, using an RBM with 10 visible units ($m=5$) and 3 hidden units.
The cut-off point was at 31 photons both for RBM and MaxLik. 
The motivation for choosing this relatively large Hilbert space is to explore the case in which the number parameters optimized by the RBM is much less than MaxLik.

Figure \ref{fig:Comparison}(a) shows the photon statistics of the state reconstructed using the two methods. 
Theoretically, we expect this state to show Poisson statistics for odd photon numbers, but zero probability for even photon numbers. 
We see that the state reconstructed using RBMs largely follows this rule whereas the MaxLik reconstructed state has significant nonzero statistics for even photon numbers. 
In Fig.~\ref{fig:Comparison}(b) we plot the fidelity of the reconstructed state with the original one as a function of the data set size and observe that RBM performs significantly better. For example, the RBM reconstructs the state from 1000 quadrature samples with the same fidelity of 98.5\% as does MaxLik from 5000 samples. This is of value because complex quantum state engineering experiments typically produce desired states at very low rates \cite{Sychev2017}, so the usage of RBM can greatly reduce the data collection effort.

The improved performance of the RBM approach for a smaller amount of experimental data is likely associated with lower overfitting~\cite{Troyer2017}. 
Indeed, the number of parameters in MaxLik is, as discussed $32^2-1=1023$, whereas for RBM it is $2\times(10\times3+10+3)=86$. 
In order to demonstrate that overfitting is indeed the cause for poorer performance of MaxLik, we implement the following cross-validation test.
We generate multiple quadrature data sets of the the same size and reconstruct the state from one of them. 
Then we calculate the log-likelihood \eqref{LogLik} for the data from each set with respect to the reconstructed state. 
If overfitting plays a significant role in the reconstruction, the likelihood of the ``native" data set (from which the state was reconstructed) is expected to be significantly higher than for other sets. 
We plot the mean difference of the log-likelihoods for the ``native" and ``non-native" data sets in Fig.~\ref{fig:Comparison}(c) and observe this difference to be much higher for MaxLik than for RBM. 
This confirms our hypothesis.

To test the generality of our conclusions, we applied RBM reconstruction to three states of different nature, Gottesman-Kitaev-Preskill~\cite{Gottesman2001}, squeezed-displaced vacuum and random states (see Supplementary). We observed the same results as for the cat state. 
This corroborates our hypothesis that the likely reason for RBM's superiority to MaxLik is that the former method is less prone to overfitting.

\begin{figure}[t!]
	\begin{center}
	\includegraphics[width=\linewidth]{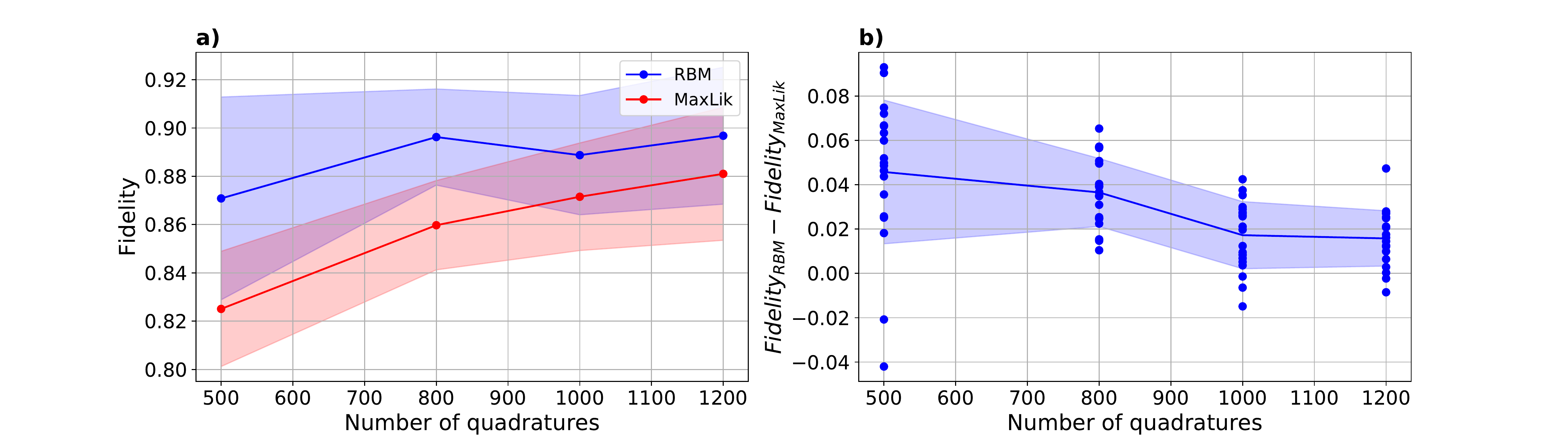}
	\end{center}
	\vskip - 6mm
	\caption{Reconstruction of random states. 
	a) Reconstruction fidelity as a function of   number of quadrature measurements. Each point is averaged over multiple states, with a single quadrature dataset generated for each state. Shaded regions show the standard deviation.
	b) Differences between RBM and MaxLik fidelities for individual random states.	}
	\label{fig:Random_states}
\end{figure}

As a further test, we applied RBM and MaxLik reconstruction to a set of 20 random superpositions of Fock states from 0 to 31 photons. 
To generate these superpositions, a Gaussian random number generator with zero mean and unit variance was used to generate the real and imaginary amplitudes of each Fock component, and the resulting states were subsequently normalised to unity. 
The performance of the reconstruction (with $N_{\rm ph}=31$) is shown in Fig.~\ref{fig:Random_states}, demonstrating the advantage of RBM-based QST for a great majority of states.

\section{Discussion and outlook}

As with any tomography method, a technique for estimating the state reconstruction precision is needed. 
The standard approach to this task is bootstrapping --- that is, generating a multiple simulated quadrature sample sets from the reconstructed state and reconstructing a state from each of these sets. 
The variance of these ``secondary" states with respect to the originally reconstructed one gives an estimate for the statistical uncertainty of the reconstruction. This approach, as well as more advanced error estimation methods~\cite{Faist2016,Wang2019}, can be successfully applied to RBM tomography.

A related question is the number of hidden units in the RBM. 
On the one hand, increasing this hyperparameter improves the reconstruction quality because of the better expressive capacity of the neural network. 
On the other hand, it increases the computation complexity. 
The compromise depends on the specific state being reconstructed. 
For example, the squeezed-displaced vacuum state, whose wave function is Gaussian and does not have multiple fine features, required less hidden units than the cat and Gottesman-Kitaev-Preskill states. 
For all states we tested, the reconstruction quality improvement saturated for the number of hidden units being similar to or less than the number of visible units.

Our results demonstrate that the neural network QST approach is a promising way of characterizing the states observed in optical experiments. 
We found this method to be capable of reliable state reconstruction and much less prone to overfitting compared to standard MaxLik approach. 
However, the full capability of our method is expected to be unveiled for very large Hilbert spaces, to which traditional methods become inapplicable. 
Therefore the natural next step would be to implement a complex multimode entangled state and apply RBM for its reconstruction. 
Promising sources of such states are multimode parametric oscillators that have seen rapid development in recent years~\cite{Furusawa2016,Fabre2017}.

As stated earlier, the complexity of the QST problem, as well as the number of required measurements, grows exponentially with the system size. 
The RBM Ansatz appears to circumvent this issue, as the number of RBM parameters is polynomial with respect to the system size. 
The price to pay is that it may not be possible to efficiently describe all states with this Ansatz. 
In the discrete-variable domain, there exists a known class of physically  interesting quantum states that carry no efficient RBM description~\cite{Gao2017}.
 It is important to undertake a similar study for continuous-variable systems to understand the application range of this method --- in particular, to which extent it can be used in the multimode case.

To proceed in the direction of large systems, we will also need to change the strategy of RBM training. 
Presently, our evaluation of the likelihood function relies on exhaustive summation of amplitudes for all elements of the Hilbert space basis (see Supplementary). 
However, such a summation will be impossible in large Hilbert spaces. 
Instead, we will have to rely on approximate methods of RBM training such as contrastive divergence~\cite{carreira2005contrastive} or Gibbs sampling~\cite{Geman1984} to select the basis elements with largest amplitudes. 
Alternative neural network architectures should also be explored.  
In particular, it would be interesting to look for ways to utilize forward propagating neural networks, rather than RBMs, for QST~\cite{Sharir2020}. 
Such neural networks are more common in modern machine learning because their training is much more straightforward. 

Our approach can be generalized to broader classes of physical problems. 
First, in addition to light, it is applicable to any physical system that can be mapped to a harmonic oscillator, such as atomic ensembles~\cite{Hammerer2010} and nanomechanics~\cite{Aspelmeyer2013}.
Second, we reiterate that neural-network based QST studied here belongs to a larger class of problems in which one looks for a quantum state that best satisfies a certain criterion. 
A particularly promising field of research, in our opinion, is complex phenomena in condensed matter systems, such as many-body localization, and describing exotic phase transitions. 
Approaches based on machine learning constitute a new and promising way of tackling them. 

{\bf Funding}.
This work is supported by Russian Science Foundation (19-71-10092).

{\bf Acknowledgements}.
We are grateful to D. Sychev for a fruitful discussion and valuable remarks. 

{\bf Disclosures}. 
The authors declare no conflicts of interest.

\newpage
\begin{widetext}
\renewcommand{\thesection}{S\arabic{section}} 
\renewcommand{\theequation}{S\arabic{equation}}
\renewcommand{\thefigure}{S\arabic{figure}}
\setcounter{equation}{0}
\setcounter{figure}{0}

\newpage\newpage
\hrulefill
\section*{Experimental quantum homodyne tomography via machine learning. Supplementary material.}

\section{Calculating gradients} 

\paragraph{Pure states.} 
Here we present the details of training our RBM. 
The neural network parametrization for the wavefunction is defined by Eqs.~(4) and (5) in the main text. 
We introduce additional notation. 
First, because there is one-to-one correspondence between visible layer configurations $\ket{\mathbf v}$ and Fock states $\ket n$, as discussed in the main text, we will use the symbol $n$ to denote both these objects. 
Second, we denote the unnormalized Boltzmann probability 
\begin{equation}
	P_n^\lambda=Z_\lambda p^{\lambda}_n=\sum_{\mathbf h} e^{-E^{\lambda}(n,{\mathbf h})},
\end{equation}
with $Z_\lambda\equiv\sum_{n}P^{\lambda}_n$, for the amplitude RBM, and the analogous quantity 
\begin{equation}
	P_n^\mu=e^{\phi_n}=\sum_{\mathbf h} e^{-E^{\mu}(n,{\mathbf h})}
\end{equation}
for the phase RBM. We remind the reader that the letters $\lambda$ and $\mu$ denote the respective parameter sets of these RBMs.

By plugging the expression~(\ref{pure}) in the main text into the log-likelihood function~(\ref{LogLik}) 
in the main text and using that the fact that overlap between the number and quadrature eigenstates $\langle \theta,X\,|n\rangle$ corresponding to the phase $\theta$ is just the $n$'s harmonic oscillator eigenfunction (Hermite-Gaussian polynomial) $H_n(X)$ with a phase factor,
\begin{equation}
	\langle \theta,X\,|n\rangle=H_n(X)\exp[-in\theta],
\end{equation}
we obtain the following expression:
\begin{eqnarray}\label{Xi}
	\Xi(\lambda,\mu)&=&\sum_j \log \left|\langle \theta_j,X_j|\,\Psi\rangle\right|^2=\sum_j \log \left|  \sum_n \langle \theta_j,X_j\,|n\rangle\langle n|\, \Psi\rangle\right|^2=-N_{m}\log Z_{\lambda}+\sum_j \log \left| \sum_n Q^j_n\right|^2,
\end{eqnarray}
where $Q^j_n=H_n(x_j)\sqrt{P_{n}^{\lambda}}\exp[-i(n\theta_j-\phi^{\mu}_n/2)]$, the summation with index $j$ is over all quadrature measurements and $N_m$ is the number of measurements.

For the network training, we evaluate the gradients of the above log-likelihood  $\Xi(\lambda,\mu)$ over the neural net parameters $\lambda$ and $\mu$ as follows:
\begin{subequations}\label{GrXi0}
\begin{eqnarray} \label{GrXi}
	&&\nabla_{\!\lambda}\Xi(\lambda,\mu)=N_{m}\{D_{\lambda}\}_{p_\lambda}-\mathrm{Re} \sum_j  \{D_{\lambda}\}_{j} \\
	\label{GrXi2}
	&&\nabla_{\!\mu}\Xi(\lambda,\mu)=  \mathrm{Im}  \sum_j  \{D_{\mu}\}_{j},
\end{eqnarray}
\end{subequations}
where we defined
\begin{subequations}\label{av}
\begin{eqnarray}
	\{D_{\lambda}\}_{p_\lambda}=\sum_{n}  D_{\lambda}^np^{\lambda}_n, \\ 
	\{D_{\mu,\lambda}\}_{j}=\frac{\sum_n D_{\mu,\lambda}^nQ^j_n}{\sum_n Q^j_n},
\end{eqnarray}	
\end{subequations}
with $D_{\mu}^n\equiv \nabla_{\!\mu}\log P^{\mu}_n$ and  $D_{\lambda}^n\equiv \nabla_{\!\lambda}\log P^{\lambda}_n$. Ascending by these gradients, we can maximize the log-likelihood \eqref{Xi}. Both RBMs are trained simultaneously.

We note that the above gradients contain exhaustive summation over possible configutations $\{n,{\mathbf{h}}\}$ of the visible and hidden layers of both RBMs. 
In the present work, we are able to compute this sum directly  since the number of RBM units is relatively small. 
However, in the case of high Hilbert space dimension, Boltzmann sampling using an annealing device or algorithm will be required. 

\paragraph{Mixed states.}

As discussed in the main text, see Eq.~(7) in the main text, we treat the mixed state $\hat\rho$ to be reconstructed as a partial state of a pure state $\ket{\Psi_{SE}}$ in a tensor product Hilbert space with the dimension $(N_{\rm ph}+1)\times (N_{\rm ph}+1)$. 
We decompose this state in the Fock basis  $\ket{\Psi_{SE}}\equiv\sum_{n=0}^{N_{\rm ph}}\sum_{m=0}^{N_{\rm ph}}C_{nm}\,|n,m\rangle$ and apply the same parametrization as in the previous subsection:
\begin{eqnarray}\label{AnsatzPhi}
	&&C^{\lambda,\mu}_{nm}=\langle n,m|\Psi_{SE}\rangle\equiv  \sqrt{p^{\lambda}_{nm}}e^{i\phi^{\mu}_{nm}/2}, \\ 
	&&\textrm{with  }Z_\lambda=\sum_{n m}P^{\lambda}_{nm}.
\end{eqnarray}
The partial trace of this state over the environment is as follows:
\begin{eqnarray} \label{AnsatzRho}
	\hat\rho^{\lambda,\mu}_{nm}=&&\sum_k C^{\lambda,\mu}_{nk}(C^{\lambda,\mu}_{mk})^*=\frac{1}{Z_\lambda}\sum_k\sqrt{P^{\lambda}_{nk}P^{\lambda}_{mk}} e^{i(\phi^{\mu}_{nk}-\phi^{\mu}_{mk})/2}.
\end{eqnarray}
The log-likelihood (Eq.~(6) in the main text) is then given by
\begin{eqnarray}\label{XiRho}
	\Xi(\lambda,\mu)&=&-\sum_j \log \sum_{nm} \langle \theta_j,x_j\,|n\rangle\,\rho^{\lambda,\mu}_{nm}\,\langle m |\theta_j, x_j\rangle \nonumber\\
	&=&N_m\log Z_\lambda-\sum_j \log   \sum_{nmk} Q^j_{nmk},
\end{eqnarray}
where the summation indices $n,m,k$ run over the truncated Fock basis, $j$ over all quadrature measurements, and
\begin{eqnarray}
	Q^j_{nmk}&&=H_n(x_j)H_m(x_j)\sqrt{P^{\lambda}_{nk}P^{\lambda}_{mk}}\times \nonumber \\
	&&\times\exp{i[(m-n)\theta_j+(\phi^{\mu}_{nk}-\phi^{\mu}_{mk})/2]}.
\end{eqnarray}
We note that the expression (\ref{XiRho}) is very similar to the pure state case (\ref{Xi}), but requires two additional summations over the truncated Fock basis.   
The log-likelihood (\ref{XiRho}) gradients over $\mu$ and $\lambda$ read similarly to those for the pure state~(\ref{GrXi0}), but with the parameters~\eqref{av} redefined as follows:
\begin{subequations}\label{avRho}
	\begin{eqnarray} 
	&&\{D_{\lambda}\}_{p_\lambda}=\sum_{nm}  D_{\lambda}^{nm}p^{\lambda}_{nm}, \\
	&&\{D_{\mu, \lambda}\}_{j}=\frac{ \sum_{nmk}D_{\mu, \lambda}^{nk}Q^j_{nmk}}{\sum_{nmk} Q^j_{nmk}}
\end{eqnarray}
\end{subequations}
The remainder of the treatment replicates that for pure states.

\paragraph{Efficiency correction.}
To correct for an imperfect  homodyne detector efficiency $\eta < 1$ in our neural net approach, we model it as a perfect detector preceded by beam splitter of transmission $\eta$ \cite{Lvovsky2004}, 
which changes the quantum state  $\hat\rho $ by means of generalized Bernoulli transformation to a new state $ \hat\rho_{\eta}$ according to 
\begin{equation}
\bra{m}\hat\rho_\eta\ket{n}=\sum_{k=0}^\infty
	B_{m+k,m}(\eta)B_{n+k,n}(\eta)\bra{m+k}\hat\rho\ket{n+k},
\end{equation}
where $B_{n+k,n}=\sqrt{\left({n+k}\atop{n}\right)\eta^n(1-\eta)^k}$.
Now we can repeat the above procedure for the mixed state (purification) Ansatz, with the only difference that we use $\hat\rho_\eta$ instead of $\hat\rho$ to calculate the log-likelihood (Eq.~(6) in the main text). 

\section{Reconstruction without loss correction}
\vspace{\baselineskip}

Figure \ref{fig: exp wigners} shows reconstruction from the experimental data corresponding to Fig.~2 in the main text, but without correction for the losses. The fidelity between the MaxLik and RBM methods is 99.6\% for the cat state, and  99.8\% for the engineered Fock superpositions. 

\begin{figure*}
	\includegraphics[width=\linewidth]{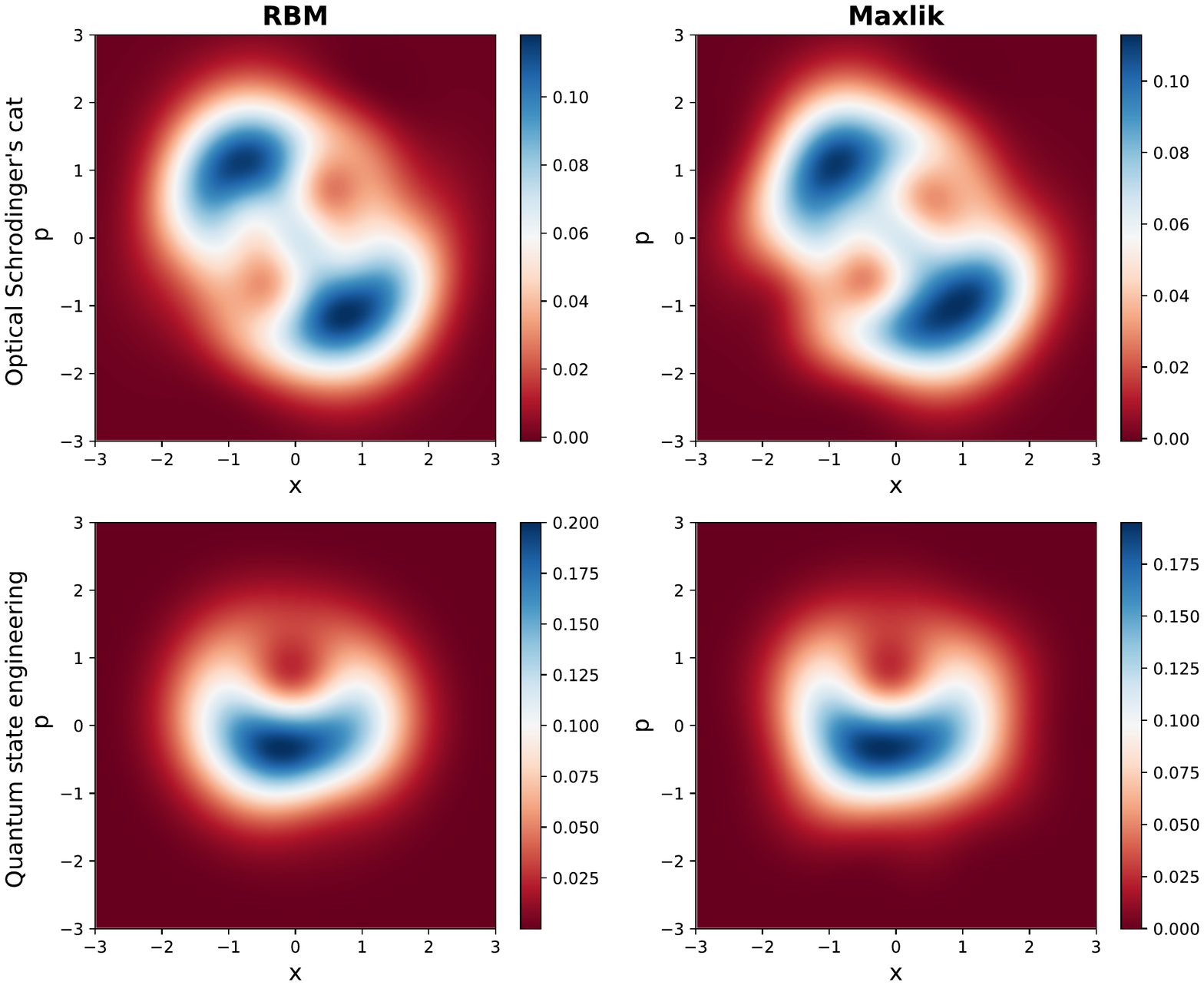}
	\vspace{-1.5cm}
	\caption{Wigner functions of the reconstructed states from the experimental data without correction. }
	\label{fig: exp wigners}
\end{figure*}

\section{Reconstruction of GKP, Squeezed-displaced and random states}

To test the generality of our conclusions, we applied RBM reconstruction to three states of different nature.
\begin{itemize}
	\item Gottesman-Kitaev-Preskill (GKP) state~\cite{Gottesman2001}, which useful for continuous-variable quantum computation and error correction protocols. This state is a superposition of equally displaced squeezed vacua with a Gaussian envelope. One can encode a qubit in GKP states as follows  
	\begin{flalign}
		\ket{\tilde{0}} &= N_0  \sum_{s=-\infty}^{+\infty} e^{-\frac{1}{2}\kappa^2(2s\alpha)^2} D(2s\alpha)\ket{0_{\xi}}\\
		\ket{\tilde{1}} &= N_1  \sum_{s=-\infty}^{+\infty} e^{-\frac{1}{2}\kappa^2[(2s+1)\alpha]^2} D[(2s+1)\alpha]\ket{0_{\xi}}
	\end{flalign} 
where $D(\alpha)=e^{\alpha a - \alpha^* a^\dag}$ is the position displacement operator, $\ket{0_{\xi}}$ is a squeezed vacuum, $\kappa$ parametrizes gaussian envelope. 
Here we have reconstructed the equal superposition of two logical bits ($\ket{\tilde{0}}+\ket{\tilde{1}}$) with parameters $\xi = 0.8, \kappa = 0.5, \alpha = 1.5$, $s$ varying from $-4$ to $4$. 
	\item Squeezed-displaced vacuum state
	\begin{equation}
		\ket{\psi(\alpha,\xi)} = e^{\xi a^2 - \xi^* a^{\dag2}} D(\alpha) \ket{0}.
	\end{equation} 
States of such a type are common in quantum optics and utilized for manipulating nonclassical light in different settings. The parameters of  the reconstructed state are $\alpha = 0.5, \; \xi = -i$.
	\item Random Fock state superposition up to 31 photons.  Gaussian random number generator with zero mean and unit variance was used to generate real and imaginary part of the state. Then the state was normalized to unity.
\end{itemize}

The reconstruction Hilbert space was cut off at $N_{\rm ph}= 31$ photons. 
The RBM for the GKP and random state had 10 hidden units, while that for the squeezed-displaced state had 1 hidden unit. 
For all these states we observed similar behavior as for the cat state both for the reconstruction fidelity and the cross-validation test (Fig.~\ref{fig:Comparison}). 

\begin{figure*}[t!]
	\begin{center}
	\includegraphics[width=1.\linewidth]{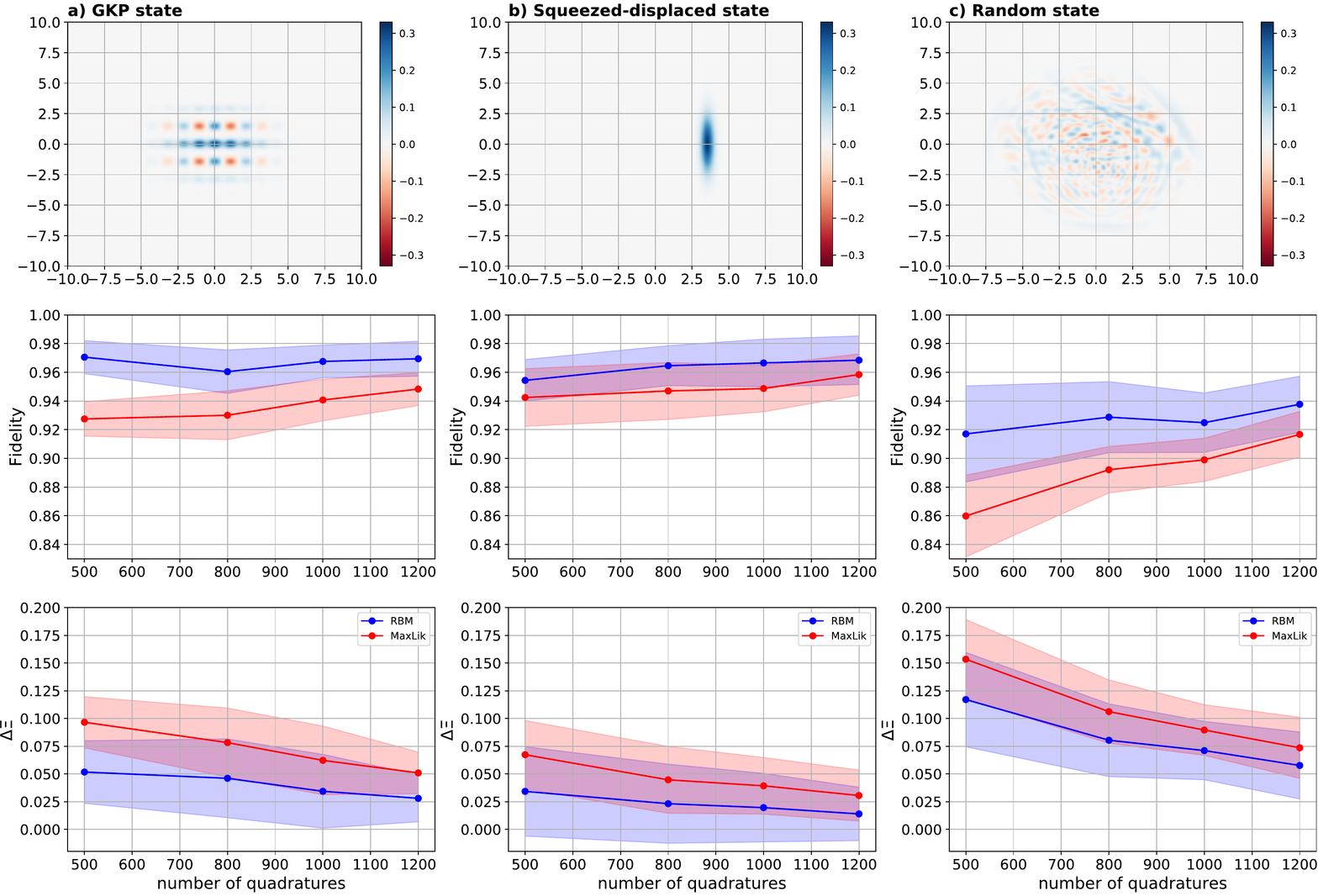}
	\end{center}
	\caption{Reconstruction of a GKP (a), squeezed-displaced (b) and random (c) states from bootstrapped data. Top: Wigner function; middle: fidelity; bottom: cross-validation log-likelihood difference. The notation is the same as in Fig.~3 in the main text.}
	\label{fig:Comparison}
\end{figure*}

\end{widetext}
\end{document}